\documentclass[conference]{IEEEtran}
\usepackage{graphicx}
\graphicspath{ {Figures/} }
\usepackage{color}
\usepackage[justification=centering]{caption}

\usepackage{cite}

\ifCLASSINFOpdf
\else
\fi

\begin{document}

\title{Real-Time Simulation and Hardware-in-the-Loop Approaches for Integrating Renewable Energy Sources into Smart Grids: Challenges \& Actions}

\author{\IEEEauthorblockN{
V.~H.~Nguyen\IEEEauthorrefmark{1},
Y.~Besanger\IEEEauthorrefmark{1}, 
Q.~T.~Tran\IEEEauthorrefmark{2}, 
T.~L.~Nguyen\IEEEauthorrefmark{1}\IEEEauthorrefmark{2},
C.~Boudinet\IEEEauthorrefmark{1},
R.~Brandl\IEEEauthorrefmark{3},
F.~Marten\IEEEauthorrefmark{3},
A.~Markou\IEEEauthorrefmark{4},\\
P.~Kotsampopoulos\IEEEauthorrefmark{4},
A.~A.~van~der~Meer\IEEEauthorrefmark{5},
E.~Guillo-Sansano\IEEEauthorrefmark{6},
G.~Lauss\IEEEauthorrefmark{7},
T.~I.~Strasser\IEEEauthorrefmark{7},
K.~Heussen\IEEEauthorrefmark{8}
}
\vspace{0.25cm}
\IEEEauthorblockA{
\IEEEauthorrefmark{1}University Grenoble Alpes, G2Elab, Grenoble, France, van-hoa.nguyen@grenoble-inp.fr}
\IEEEauthorblockA{
\IEEEauthorrefmark{2}Commissariat \`{a} l'\'{e}nergie atomique et aux \'{e}nergies alternatives, Chamb\'{e}ry, France}
\IEEEauthorblockA{
\IEEEauthorrefmark{3}Fraunhofer Institute of Wind Energy and Energy System Technology, Kassel, Germany}
\IEEEauthorblockA{
\IEEEauthorrefmark{4}National Technical University of Athens, Athens, Greece}
\IEEEauthorblockA{
\IEEEauthorrefmark{5}Delft University of Technology, Delft, The Netherlands}
\IEEEauthorblockA{
\IEEEauthorrefmark{6}University of Strathclyde, Glasgow, United Kingdom}
\IEEEauthorblockA{
\IEEEauthorrefmark{7}AIT Austrian Institute of Technology, Vienna, Austria}
\IEEEauthorblockA{
\IEEEauthorrefmark{8}Danmarks Tekniske Universitet, Kongens Lyngby, Denmark}
}

\maketitle

\begin{abstract}
The integration of distributed renewable energy sources and the multi-domain behaviours inside the cyber-physical energy system (smart grids) draws up major challenges. Their validation and roll out requires careful assessment, in term of modelling, simulation and testing. The traditional approach focusing on a particular object, actual hardware or a detailed model, while drastically simplifying the remainder of the system under test, is no longer sufficient. Real-time simulation and Hardware-in-the-Loop (HIL) techniques emerge as indispensable tools for validating the behaviour of renewable sources as well as their impact/interaction to with the cyber-physical energy system. This paper aims to provide an overview of the present status-quo of real-time and HIL approaches used for smart grids and their readiness for cyber-physical experiments. We investigate the current limitations of HIL techniques and point out necessary future developments. Subsequently, the paper highlights challenges that need specific attention as well as ongoing actions and further research directions.
\end{abstract}



\section{Introduction} 
The decarbonization scenario of the European power generation requires a high penetration of Distributed Renewable Energy Sources (DRES). The expected high proportion of DRES integration and limited storage capabilities give rise to new challenges to power system operators in maintaining the security of supply and the power quality, such as the fulfillment of the established voltage quality standards. The intermittent behaviour of DRES also leads to new requirements in testing and validating innovative solutions in the integration of such devices into the power systems. As considered by the IEEE Power and Energy Society Task Force on ``Real-Time Simulation of Power and Energy Systems'', real-time simulation, especially the use of Hardware-in-the-Loop (HIL) approaches can fulfill several requirements in dealing with power system stability assessment and rapid prototyping to ensure and validate safety and security of power system operation with innovative solutions. In this approach, a real hardware setup for a domain (or part of a domain) is coupled with a simulation tool to allow testing of hardware or software components under realistic conditions. The execution of the simulator in that case requires strictly small time steps in accordance to the real-time constraints of the physical target. On the other hand, HIL provides the possibility of replacing inaccurate or incomplete models with real-world counterparts.

Some of the main advantages of HIL are the thorough study of transient and steady state operation of a Hardware-Under-Test (HUT) under realistic, yet safe and repeatable, conditions; testing of a HUT in faulted and extreme conditions without damaging laboratory equipment; maintaining flexibility in choosing test parameters and components; as well as maintenance of a relatively safe environment for the personnel. In the domain of smart grid validation, HIL techniques have been successfully used for a wide range of experiments, from single devices validation, such as: protection relays \cite{Almas2012}, power system devices \cite{Loddick2011}, novel power converter structures \cite{Kotsapopoulos2012}, 
to whole cyber-physical energy systems \cite{Bian2015,Nguyen_Multi_2017}.

While offering a wide range of possibilities for smart grid validation, particularly the integration of DRES and the multi-domain behaviours of cyber-physical energy systems, HIL technique remains young and suffers from a few limitations, inter alia, the difficulty in integration of HIL to the communication layer, the limited capacity of HIL simulation for complex system and remote HIL or distributed HIL for joint experiments. Moreover, due to the immature standardization process of HIL techniques, information exchange occurs via different proprietary interfaces and there is no general framework to support the reusability of experiments.

In this paper, we cover the status-quo of real-time and HIL technologies, their present limitations in terms of experiments concerning integration and impact of DRES into the cyber-physical energy system. The technical challenges along with the current actions of the research community to resolve those limitations are investigated. 

The paper is organized as follows: Section~\ref{sec:hil_technique} provides an overview of the state-of-the-art of HIL approaches. Section~\ref{sec:difficulties} outlines the main challenges of future developments. A discussion about necessary future research and technology developments is provided in Section~\ref{sec:future_directions}.

\section{Current limitations of HIL techniques and necessary developments}
\label{sec:hil_technique}

\subsection{HIL techniques applied in the context of smart grids}

The usage of HIL techniques in smart grid applications is generally classified into Controller Hardware-in-the-loop (CHIL) and Power-Hardware-in-the-loop (PHIL) \cite{Faruque:2015,Steurer:2007}. CHIL involves the testing of a device, for example a power converter controller, where signals are exchanged between a real-time simulator and the HUT via its information ports. The interface in that case (CHIL) consists of Analogue to Digital and Digital to Analogue converters and/or digital communication interfaces. Besides control devices as HUT although real-time simulations coupled to other units such as relays, Phasor Measurement Units (PMU) or monitoring components are usually classified as CHIL. Such devices are validated in a closed-loop environment under different dynamic and fault conditions, therefore enhancing the validation of control and protection systems for power systems and energy components. In contrast, PHIL involves the testing of a device which absorbs or generates power (e.g. Photovoltaic inverter). A power interface is therefore necessary. 

\begin{figure}[!htbp]
    \begin{center}
    \includegraphics[width=0.45\textwidth]{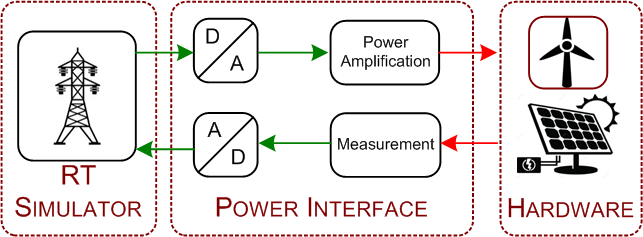}
    \caption{Basic elements of a PHIL experiment.}
    \label{fig:PHIL_Setup}
    \end{center}
\end{figure}

A general HIL setup consists of three main elements, \textit{(i)} the Real-Time (RT) simulator, \textit{(ii)} the HUT, and \textit{(iii)} the power interface (only in PHIL case) as depicted it Fig.~\ref{fig:PHIL_Setup} \cite{Faruque:2015,Steurer:2007}.

\begin{paragraph}{The RT simulator} Computes the simulation model in real-time and offers I/O capacities to reproduce the behavior of the simulated system under dynamic conditions. The simulator allows designing and performing various test scenarios with a great flexibility. Most real-time simulators have multiple processors operate in parallel to form the target platform on which the simulation runs in real-time and I/O terminals to interface with external hardware. A host computer is used to prepare the model off-line and then compile and load it on the target platform. Host computers are also used for monitoring the results of real-time simulation. A communication network exchanges data between multiple targets when the model is split into multiple subsystems in which a separate communication link is reserved for data exchange between the ``Host'' and the ``Target''. 

\end{paragraph}

\begin{paragraph} {The Hardware-Under-Test} CHIL allows testing of physical controller devices, such as DRES controllers, relays, PMU, etc., while PHIL involves also a wide variety of DRES devices and networks such as converter devices, electric vehicles and corresponding charging equipment or whole micro-grids can be test in realistic environments.
\end{paragraph}

\begin{paragraph}{Power interface} A power interface in PHIL experiments generally consists of a power amplifier and sensors that transmit measurements in feedback. It allows the interaction of the virtual simulated systems with the HUT.
\end{paragraph}

In cyber-physical energy systems, the closed loop behaviour of non-conventional units at slower time scales, such as for example battery and PV systems in a demand response context, is also of research interest. To scale up the relevant behaviours for investigation and testing, also such units would ideally be integrated into the simulated world in a ``soft''-HIL manner and not necessarily real-time. Co-simulation \cite{Palensky2016} and ``soft'' HIL setups (e.g. \cite{Kosek2013, Kullmann2011, LeRay2016, Thavlov2016}) have proven effective to demonstrate and evaluate closed-loop effects in such setups. It has remained difficult to bridge the two worlds of ``hard'' PHIL/CHIL and this ``soft'' HIL and co-simulation in a systematic framework. 

\subsection{Limitations of HIL techniques}

Offering a wide range of possibilities for validation and testing of smart grid systems, current HIL technology still has several limitations:

\begin{itemize}
    \item Limited capacity of HIL simulation for complex systems (e.g. scale effects, synchronicity, diversification), for studies of non-linearity, high harmonics and transient phases. Sometimes the fidelity to represent the dynamics of complex power components such as power electronics converters have to be compromised due to the fixed time-step of real-time simulations, limiting the size of the simulations and the transient performance.
	\item Limited capacity of remote HIL and geographically distributed HIL for joint experiments, mostly due to synchronization (CHIL and PHIL) and power interface stability and accuracy with respect to loop delay (PHIL).
	\item Difficulty in integration of HIL technology to the communication layer, particularly related to the synchronization of real-time and offline simulation, as well as continuous and discrete timelines. The communication network in simulation or in a lab does not always reflect the real scenario where the long geographic distance among different equipment may cause unexpected delay and signal loss and may cause failure for timely control. Until now, the communication network is usually separately simulated with dedicated software in order to study the effect of realistic latency, packet loss or failure in the Information and Communication Technology (ICT) system on the reliability and performance of monitoring and control applications and to what extent we have to invest on ICT infrastructure to satisfy the requirements. Communication simulators facilitate also cyber-security related experiments, such as denial-of-service protection, confidentiality and integrity testing, which is important but not always easy to the electrical community. A holistic consideration of the cyber-physical energy system, as well as impact of ICT issue to the power system requires therefore a seamless integration of HIL technique to the co-simulation framework.
	\item Lack of a general framework to facilitate the reusability of models, information exchange among different proprietary interfaces or among different partners of a joint HIL experiment.
\end{itemize}

Due to the aforementioned issues, several aspects of HIL technology need to be improved in the near future to adapt to the rapidly evolving integration of DRES to the grid. Overall, the following research trends can be observed and addressed:

\begin{paragraph}{Integration of co-simulation and HIL}
Although connection architectures may vary via means of ad-hoc connections or in a master/slave fashion, a co-simulation framework allows in general the joint and simultaneous simulation of models developed with different tools, in which the intermediate results are exchanged during simulation execution. In the domain of smart grid nowadays, the co-simulation approach is often used to couple a power system simulator and a communication simulator. It is expected that integrating HIL technology into co-simulation frameworks is an important contribution toward a holistic approach for experimenting with cyber-physical energy systems \cite{Nguyen_Cosim_2017}. Combining the strengths of both approaches, multi-domain experiments can be studied with realistic behaviours from hardware equipment under a variety of complex environments, co-simulated by appropriate and adapted simulators from the relevant domains. It will enable a complete consideration of the electrical grid to be interconnected with other domains. Most of the existing work involving integration of HIL and co-simulation uses only a direct coupling with the real-time simulator \cite{Bian2015} or a CHIL setup \cite{Rotger2016}. 
\end{paragraph}

\begin{paragraph}{Remote and geographically distributed HIL}
Latency strongly influences the accuracy (HIL) and  the stability (PHIL) of a HIL test. Moreover, random packet loss due to network congestion outside of a Local Area Network (LAN) may alter the information and cause malfunction at the real-time simulator, including any connected hardware \cite{Liu2017}. Up-to-now scientists have investigated the possibility of extending PHIL beyond laboratory geographical boundaries, and mostly, for latency tolerant applications (e.g., monitoring) \cite{Lundstrom2016}. These developments could be a first step in enabling the possibility of remote HIL and geographically distributed HIL.
\end{paragraph}

\begin{paragraph}{Interoperability and standardization} 
Within a HIL co-simulation test it is crucial to ensure seamless communication among the individual components and simulators. Additionally, when the experiments involve multiple domains or multi-laboratories, it is required to have strong interoperability between different partners \cite{Nguyen2016}. A common information model is necessary to enable seamless and meaningful communication among applications. First attempts have been made towards creating a common reference model to improve interoperability and reusability of HIL experiments \cite{Andren2014}. With these efforts towards harmonization and standardization of HIL technology, a standardized and general framework for HIL experiments can be established.
\end{paragraph}

\section{Major technical challenges}
\label{sec:difficulties}

Major technical challenges towards resolving the aforementioned limitations are as follows.

\subsection{Data flow and concurrency}

Within the process of a HIL experiment and particularly the integration of HIL to co-simulation, it is crucial to ensure a synchronous data flow among the individual components, as well as the concurrency of the simulators. While strongly influencing the accuracy of the experiment result, ensuring this synchronization and concurrency is not always an easy task, especially in the case of real-time and HIL simulation and when it is necessary to take into account the impact of ICT. Synchronous data flows are required in all the software/software and software/hardware interface. The synchronization of an non-real-time simulation with a real-time simulation as well as the harmonization of their time steps, are not always evident. 

On the other hand, when it involves the co-simulation of power system and communication network for an integrated analysis of both domains. It is necessary to synchronize both simulation tools properly at runtime. However, the existing simulation tools offer limited options of adequate Application Programming Interface (API) for external coupling. On top of that, the fundamentally different concept behind power system and communication networks is also a challenge to detect, link, and handle related events in both domains.

\begin{itemize}
    \item Power system simulation is normally continuous with possibility of event detection associated to value crossing a certain threshold. 
    \item Communication network simulation is based on discrete events whose occurrence usually unevenly distributed with respect to time. The simulator provides an event scheduler to record current system time and process the events in an event list.
\end{itemize}

When the experiments involve multiple domains or multi-laboratory, it is required to have a certain degree of interoperability among the different actors as well as among different elements of the experiments. A common information model or at least a conversion interface is necessary. In a power system simulation, the exact and proper representation of a system’s topology is critical, proportionally with scale and complexity. The employed information model should be capable to represent, encapsulate and exchange static and dynamic data, as well as, to inform any modification in topology and current state of the network in real-time and in a standardized way. Moreover, it should have capacity to include ICT aspect, which is not readily covered in existing models.

To conclude, data flow and concurrency is an obligation in HIl technique and in the integration of HIL to co-simulation framework. The two main difficulties are the synchronization of offline/real-time simulations, continuous/discrete timescales and the limited capacity of actual information model.

\subsection{Instability and reduced accuracy}

The main difficulty towards integration of HIL into a holistic framework is, inter alia, the issue of signal latency, compensation of loop delay and time synchronization. Due to the addition of the hardware, the loop time is introduced and negatively influences the synchronization algorithm. Especially for the case of PHIL experiment, the power interface, due to various external disturbance (but most importantly the loop delay), is very sensitive in term of stability and accuracy. Configuration and impact of the power amplifier (I/O boundaries, galvanic isolation, short circuit behavior, slew-rate, etc.) must be addressed and evaluated to match the specific requirement for each PHIL setup as it strongly influence the determination of system stability, bandwidth, and the expected accuracy. In order to improve the stability and accuracy of a PHIL experiment, the challenge is to synchronize and compensate the loop delay. The first step should be selection of appropriate interface algorithms and power amplification. Secondly, a time delay compensation method could be considered, such as introducing phase shifting, low-pass filter to the feedback signal \cite{Kotsapopoulos2012}, extrapolation prediction to compensate for time delays \cite{Ren2011}, phase advance calibration\cite{Roscoe2010} or multi-rate real-time simulation \cite{Viehweider2011}.

\section{Challenging the status quo}
\label{sec:future_directions}

We present in this section, the ongoing actions, addressed by the European ``ERIGrid'' project to improve the capacity of HIL experiments, in term of stability and accuracy, as well as our approaches to integration HIL into cos-imulation framework. These developments resolve the current limitation of straight-forward laboratory or pure software-based technologies and contribute to establishment of a general and holistic framework for smart grid validation.

\subsection{Integration of HIL and co-simulation framework}

As presented in the above sections, the integration of HIL to co-simulation framework allows us to have a complete view of the behavior of both domains and the physical energy system states, while power system and communication network are simulated with the suitable solver and the calculation loads are shared. Moreover, addressing holistic testing of global power system scenarios, limiting laboratory-based and/or simulation-based technologies challenging individual research infrastructures. This approach of integrating HIL to co-simulation provides the technical base and paving the way to international collaboration by combining several infrastructures and/or replacing non-available components/systems by simulation, increasing the realism of innovative validation methods.

In order to resolve the aforementioned technical challenges and ensuring synchronization and concurrency of simulators, three methods were proposed:

\begin{paragraph}{Offline integration with Functional Mock-up Unit}
In order to improve interoperability and reusability of the models developed in co-simulation frameworks. Two major standards have been issued; the Functional Mockup Interface (FMI) and the High Level Architecture framework (HLA), with FMI oriented towards model exchange and coupling simulators for co-simulation. The basic component of FMI is the functional Mock-up Unit (FMU), which gives access to the equations of the model (mode ``model exchange'') or implements a solver manipulating the equations of the model (mode ``co-simulation''). In this approach, the non-real-time simulation model is converted to FMU and integrated directly to the real-time simulator’s model and is forced to run at real-time simulators time steps. This requires however computability verification and the master algorithm in that case is the real-time simulator itself.
\end{paragraph}

\begin{paragraph}{Online integration without signal synchronization -- LabLink architecture}
Inspiring by the underlying idea of governing exchanged signals via a common ``hub'', the LabLink architecture\footnote{Developed by AIT Austrian Institute of Technology.}, however acts like a message router and does not consider the synchronization aspects (see Fig.~\ref{fig:lablink}). The proposed architecture aims not only to integrate HIL to co-simulation framework but also to link various research infrastructures in a joint experiment, hence the name. This approach provides a holistic simulation environment, which combine different solvers to their appropriate models (steady state solvers for large scale simulation and real-time capable solvers for transient real-time simulation). 

\begin{figure}[!htbp]
    \begin{center}
    \includegraphics[width=0.50\textwidth]{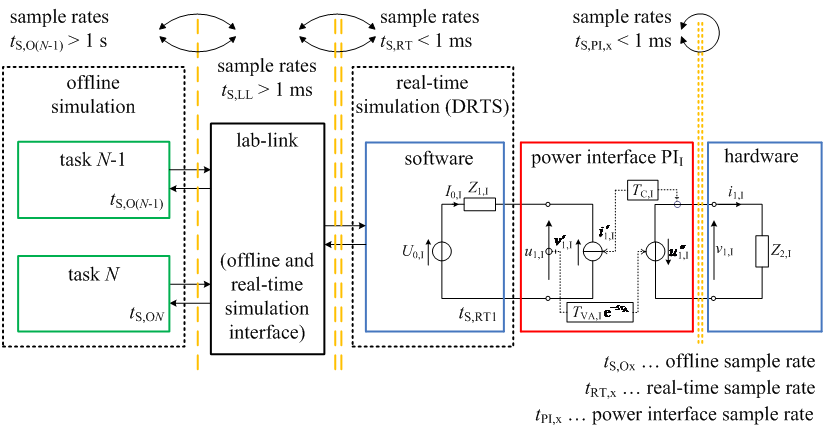}
    \caption{LabLink architecture for PHIL with recommended sample rates.}
    \label{fig:lablink}
    \end{center}
\end{figure}
\end{paragraph}

\begin{paragraph}{Online Integration with synchronization -- OpSim solution} 
Another framework which could potentially merge HIL and co-simulation is the OpSim test- and simulation-environment\footnote{Developed by Fraunhofer IWES and University of Kassel, funded by the German Federal Ministry for Economic Affairs and Energy.} (see Fig.~\ref{fig:OpSim}) which provides a client/proxy architecture with a central message bus. The environment copes with the challenge of combining several simulations and controllers to an overall holistic testbed. It allows integrating software from external partners via either a web-based remote interface or standard interfaces such as VHPready, CIM and IEC61850. 

The synchronization aspect of OpSim follows the conservative approach, which allows a simulator to proceed only for a time period in which it can be proven that no events sent out from other simulators are expected. This ensures causally correct and reproducible simulation results. The environment may be extended with physical laboratory-based domains, to increase setup realism, extend hardware validations methods (HIL capability) and combine various simulation tools by real-time capable connections due to online tool translation. For such HIL-applications, one may either use the Opal-RT system, which is connected to OpSim via its asynchronous TCP-interface, or use one of the available standard interfaces.

\begin{figure}
    \begin{center}
    \includegraphics[width=0.5\textwidth]{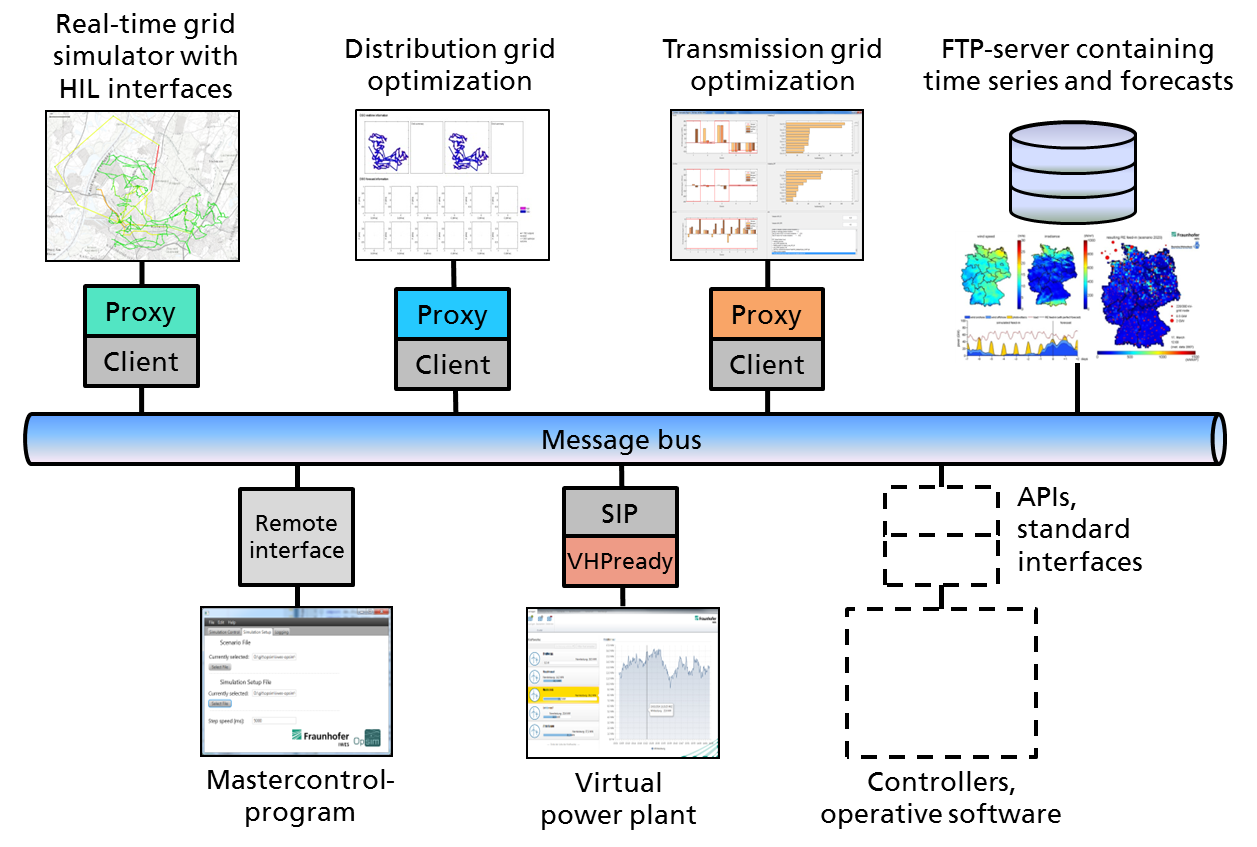}
    \caption{Possible integration of HIL and co-simulation with the OpSim-environment (image is based on the illustration on www.opsim.net)}
    \label{fig:OpSim}
    \end{center}
\end{figure}
\end{paragraph}

\subsection{Improvement to PHIL capacity}

\begin{paragraph}{Stability analysis}
It is common practice to represent PHIL equipment in the domain of frequency using Laplace transformation. However, frequency domain is not capable of describing nonlinear models. As a result, linear approximations of them are required, which, in practice means less accuracy in the analysis of PHIL simulation. In the framework of this paper, an analytical approach, extensively described in \cite{Markou2017}, is briefly presented. The model of a PHIL simulation can be express using transfer function in the frequency domain as shown in Fig.~\ref{fig:General representation}.

\begin{figure}[!htbp]
    \begin{center}
    \includegraphics[width=0.45\textwidth]{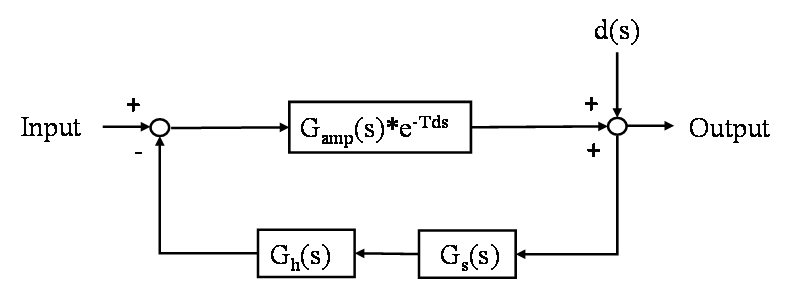}
    \caption{General representation of a PHIL system}
    \label{fig:General representation}
    \end{center}
\end{figure}

$G_s(s)$, $G_{amp}(s)$ and $G_h(s)$ are the transfer functions of simulated part, amplifier and hardware part respectively and the exponential term is the representation in the frequency domain of the time delay inserted by the amplifier. The disturbance inserted into the system due to extrinsic factors is symbolized as d(s).

Based on the previous analysis for a PHIL simulation and using Bode stability criterion the stability conditions can be expressed as: $|G_s(s)G_{amp}(s)e^{sT_d}G_h(s)|<1$ and $\angle{G_s(s)}+\angle{G_{amp}(s)}+\angle{G_h(s)}-\omega{T_d}=pi$\\
Taking into account uncertainties that occur in different parts of the model of a PHIL simulation the previous inequality related with magnitude of the open transfer function is given by: $$|G_s(s)G_{amp}(s)e^{sT_d}G_h(s)|<\frac{1}{1+\epsilon}$$
As the parameter $\epsilon$ is defined always bigger than zero then the value of the fraction of the right part of the inequality is smaller than unity. Thus, one can conclude that when there are unmodelled parts in the system intentionally or not, then the stability criterion of the analysis should be more strict. From a practical point of view, we apply a more conservative method in order to examine the bounds of the stability of the system and derive safe results even in the worst case scenario.

The proposed stability analysis were applied in two different stability methods, the feedback filter and the shifting impedance method and experimentally validated in PHIL setup as outlined in \cite{Markou2017}.
\end{paragraph}

\begin{paragraph}{Time delay compensation}

The time delay introduced in PHIL simulations affects mainly to the phase relationship between current and voltage at the point of common coupling between the simulation and the hardware under test. Accordingly, the power factor and reactive power consumption of the hardware under test is inaccurate \cite{Guillo2015}. Even if a time delay compensation method is implemented, a decrease of the overall time delay is suggested as it will improve the dynamics of the simulation. Therefore, for scenarios where fast transients are expected and to be captured and analyzed, the accumulated time delay (even if this has been compensated in some manner) will dictate the maximum accurate transient to be captured if the power interface can also perform it. For testing scenarios that require evaluation of harmonics components, phases lag suffered by the harmonic components will be larger than the fundamental and sometimes not even proportional.

Therefore, in order to precisely compensate for the inherent time delay introduced in PHIL simulations, a method based on a phase-shift of the reference signal (sent to the power interface for amplification) harmonic-by-harmonic and phase-by-phase can be used within the control of the power interface in order to avoid additional delays by using an additional device for the compensation algorithm. In this manner the time delay compensation will not affect to the system topology and therefore the dynamic behaviour of the original system will stay as it originally was in terms of power angles and V-I phase relationships for all the harmonics processed \cite{Guillo2015}. 
\end{paragraph}

\section{Conclusion}
\label{sec:conclusion}

In this paper, we presented a survey on the status-quo of real-time and HIL approaches used for validating smart grid systems and components, with respect to integration of DRES and impact to the cyber-physical energy system. The current limitations of the technology is studied and necessary developments are pointed out; the integration of HIL technique to co-simulation framework and the improvement of HIL capacity in term of stability and accuracy. Subsequently, the paper highlighted the current initiatives to pave the way towards these developments.

As also pointed out in the paper, a general framework for the validation of HIL experiments is required, to improve interoperability, reusability of HIL models and to bridge the gap between pure software simulations and hardware experiments, etc. The expected framework should provide the research community and industry with the appropriate tools (real-time, CHIL/PHIL, co-simulation) so as to derive useful results in each step of this procedure and understand the factors that affect the phenomena under investigation separately. The work presented here would be the first bricks towards such framework.


\section*{Acknowledgment}
This work is supported by the European Community's Horizon 2020 Program (H2020/2014-2020), under project ``ERIGrid'' (Grant Agreement No. 654113). 

\ifCLASSOPTIONcaptionsoff
  \newpage
\fi

\IEEEtriggeratref{8}

\bibliographystyle{IEEEtran}
\bibliography{literature}

\end{document}